\documentclass[prd,notitlepage,nofootinbib,preprintnumbers,amssymb,superscriptaddress,twocolumn]{revtex4-1}
\usepackage{amsfonts,amssymb,amsmath,graphicx,color,bm,ulem}
\usepackage{latexsym,subfigure,ascmac}
\usepackage{comment}
\definecolor{ultramarine}{rgb}{0.07, 0.04, 0.56}
\definecolor{cadmiumgreen}{rgb}{0.0, 0.42, 0.24}
\definecolor{indigo(dye)}{rgb}{0.0, 0.25, 0.42}
\usepackage[linktocpage=true]{hyperref}
\hypersetup{
colorlinks=true,
citecolor=ultramarine,
linkcolor=cadmiumgreen,
urlcolor=indigo(dye),
}

\newcommand{\be}{\begin{equation}}  
\newcommand{\ee}{\end{equation}}

\begin{document}

\title{
Physical effects of gravitational waves: pedagogical examples
}

\author{Hayato Motohashi}
\affiliation{Division of Liberal Arts, Kogakuin University, 2665-1 Nakano-machi, Hachioji, Tokyo 192-0015, Japan}

\author{Teruaki Suyama}
\affiliation{Department of Physics, Tokyo Institute of Technology, 2-12-1 Ookayama, Meguro-ku, Tokyo 152-8551, Japan}

\begin{abstract}
General relativity describes gravitation in terms of the geometry of spacetime.  
It predicts the existence of gravitational waves (GWs) that stretch and compress spacetime 
and were detected recently by state-of-the-art interferometer observations.
Yet, for those who are not familiar with general relativity,
it may be difficult to understand how the GWs actually stretch and compress spacetime.
In this paper, after reviewing the fact that the effects of GWs can be understood 
as a force in Newtonian mechanics,
we consider extreme and readily perceivable
situations where
large-amplitude GWs pass through the human body and the Earth, and demonstrate
that GWs cause phenomena that commonly occur in daily life
by back-of-the-envelope calculations.
Our analysis provides intellectual and pedagogical materials for understanding 
the nature of GWs for nonexperts in general relativity.
\end{abstract}

\maketitle  


\section{Introduction}

On February 11, 2016, the LIGO Scientific Collaboration and
Virgo Collaboration announced the first direct detection of gravitational waves (GWs) 
generated by the coalescence of binary black holes (BHs)~\cite{Abbott:2016blz,LIGOpress}.
It was a historical milestone for mankind to finally prove the prediction of Einstein's general relativity proposed about 100 years ago.
GWs are ripples in spacetime.
Now, it is almost common knowledge not only among scientists but also the general public that 
spacetime is stretched and compressed upon the passage of GWs.  
Many people are interested in GWs and general relativity; hence, it is time to think about how to cultivate its intuitive understanding~\cite{Schutz:2021xns}.
For a good review on the history of GWs, see \cite{Cervantes-Cota:2016zjc}.

\begin{figure}[t]
 \begin{center}
   \includegraphics[clip,width=0.99\columnwidth]{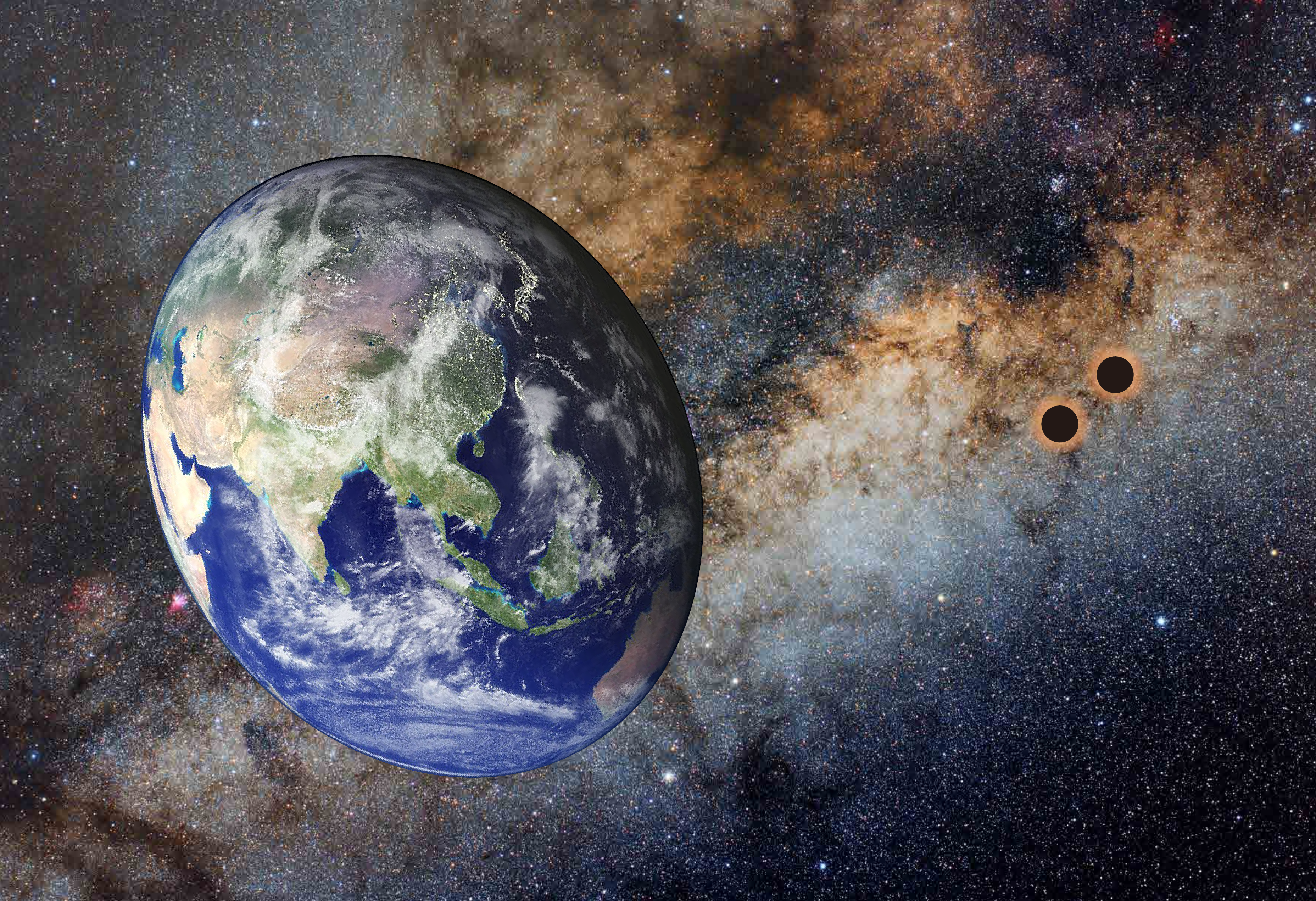}
   \caption{ 
    Artistic image of extremely exaggerated physical effects of GWs.
    Owing to the passage of GWs, the Earth is stretched and compressed in a direction perpendicular to the propagation of GWs.
    In addition to this effect, two effects on the light arriving at the observer, i.e., color change of the light due to the red-/blue-shift and the change in light trajectories, also distort the appearance
    of the Earth.}
   \label{Earth-GW}
 \end{center}
\end{figure}

Compared with, for example, electromagnetic waves, 
there are several obstacles to people being aware of the effects of GWs in their daily lives. 
First of all, it is difficult to imagine what happens when GWs stretch and compress spacetime. 
In particular, a common misconception is as follows: 
``Rulers should also be stretched and compressed similarly to space. 
Thus, are GWs not physically observable inherently?''
Actually this is not the case, and GWs do lead to physically observable effects.
In principle, one can notice the passage of GWs if their amplitude is sufficiently large 
to cause visible distortions (see Fig.~\ref{Earth-GW}).
Although this conceptual issue has already been addressed in depth in the literature (e.g., see \cite{Misner:1973prb, Creighton, Maggiore:2007ulw, Saulson:1997ck}), in the next section,
we briefly provide the physical explanation of the reason why GWs can yield observable effects,
which is basically a restatement of what is already known.
Nevertheless, we believe that it helps readers confronted with this conceptual question
recognize the reality of GWs and easily understand the subsequent sections
where we study how the GWs with extreme amplitudes cause visible effects on daily phenomena.

The second reason why people do not commonly think about GWs as much as they
do about electromagnetic waves
is their extremely small amplitude.
For instance, in the case of GWs emitted from a typical binary BH merger at a cosmological distance, 
the change in distance between the Earth and the Sun is merely about one-tenth of the Bohr radius!
It is this smallness of GW amplitude that had prevented direct detection for about 100 years and
rendered GWs irrelevant to daily physical phenomena.

Although these peculiar circumstances have led to the general public being less aware of GWs,
given their scientific significance, 
it is still regrettable if only the experts of general relativity can appreciate their physical properties\footnote{There are several educational papers that aim to
spread the basic concept and recent experimental achievements of general relativity 
and GWs \cite{Schutz:1984nf, Farr:2011cw, Burko:2016vnu, Mathur:2016cox, Hilborn:2017liy}. }.
Fortunately, the physical effects of GWs can be understood through the familiar Newtonian mechanics \cite{Misner:1973prb, Maggiore:2007ulw}. 
The aim of this paper is to provide intellectual and pedagogical materials for 
understanding the nature of GWs within the framework of Newtonian mechanics
and at the level of the back-of-the-envelope calculation, by which students who have learned 
Newtonian mechanics but are not familiar with relativity can obtain a clear view of the physical effects of GWs.
To this end, we consider extreme and 
readily perceivable situations.
As it is often the case when learning a particular subject in physics,
ideal examples enable us to easily understand the essence of the physical effects 
of GWs without being concerned with non-GW effects that are unnecessary for
the current purpose but may become dominant in more realistic situations.
We perform a quantitative analysis of several parameter regions in the amplitude--frequency space
in which GWs would visibly affect us.
Additionally, considering a BH binary system as the source of GWs, 
we shall estimate the corresponding BH mass and distance from the Earth.

We stress that our results should {\it not} be regarded as a warning of the risk
of damage to human beings posed by GWs in actual daily life.
As we shall see below, GWs may readily and perceivably exert their effects only at 
a sufficiently close distance from massive BHs.

\section{Physical effects of GWs: general arguments}
\label{sec:GA}
Before presenting our clear examples of the physical
effects of GWs, in this section we provide a general argument
on how we can, in principle, detect GWs by addressing the following conceptual question
that people who are not familiar with general relativity might ask: 
``If the passage of GWs causes nothing but the stretching and the compression of space, do they also
stretch and compress a ruler for measuring the change in the size of the body simultaneously,
making the effects of GWs unobservable?''
Although the answer to this question is already provided in several reports (e.g., \cite{Misner:1973prb, Creighton, Maggiore:2007ulw, Saulson:1997ck}),
we consider it useful not to skip this issue but to recapitulate the essential point of the answer here 
in order for readers to understand the background and hence 
the subsequent results of our analyses.

In a nutshell, an essential point in any experiment to detect the physical effects 
of GWs is the existence of an ``absolute ruler''.
Here, the absolute ruler refers to the ruler (not any ordinary ruler) 
that does not change upon the passage of GWs.
The effects of GWs can then be perceived as the change of the system measured using the absolute ruler.

The absolute ruler is a set of fundamental physical constants that are literally absolute
and do not change in the presence of GWs.
The relevant physical constants depend on the types of system and measurement considered.
Here, we focus on the following two cases of physical phenomena caused by GWs and clarify 
how GWs induce physical effects that can be detected.
\begin{enumerate}
    \item Variation of light travel time 
    \item Variation of the size of a deformable or semirigid body with time
\end{enumerate}
The relevant physical constants in each phenomenon are
the speed of light $c$ (case 1) 
and those used to determine the electromagnetic force at the microscopic level, 
such as the electric charge (case 2).

Case 1 is actually implemented in current laser interferometers (such as LIGO).
In laser interferometers, the distance $L$ between the beam splitter and the mirror changes
upon the passage of GWs, and consequently the travel time of the laser light over that distance,
which the detector actually measures, also changes.
Although the propagation speed of the laser light is not affected by GWs, 
the wavelength $\lambda$ is stretched and compressed by them;
hence, the laser light is red-/blue-shifted.
Thus, the ratio $L/\lambda$ remains constant. 
Nevertheless, this does not mean that the laser interferometer cannot detect GW signals
since it can detect the change in light travel time~\cite{Saulson:1997ck, Farr:2011cw}.

In case 2, the physical phenomena 
induced by the distortion of a deformable body or vibrations of a semirigid body are measured.
To understand this qualitatively, 
let us model the material of a body as a collection of point masses 
connected by springs. 
This model provides a convenient alternative to the electromagnetic force acting
between atoms.
The properties of the springs, such as their natural length and restoring force, are
determined by physical constants, such as the electric charge and electron mass, 
and are not affected by GWs.
When GWs pass through, 
the GWs change the length of the springs from their natural length.
For the atoms and springs, this action is perceived as a {\it physical} force
because natural length is absolute.
As a result, the atoms oscillate around 
the equilibrium position,
which, at the macroscopic level, can be seen as vibrations of the body.

For case 2, we can derive the effective force induced by GWs without resorting to general relativity.
Consider a system consisting of two point masses floating in space, each with mass $m$ and separated by a distance $L$.
When a gravitational wave with amplitude $h(t)$ passes through the system, the distance between the masses (as measured by an absolute ruler) becomes $\sqrt{1 + h(t)} L$ according to general relativity. 
If we assume that $h(t) \ll 1$, which is valid in most cases explored in this paper, we can use the binomial approximation $(1 + \epsilon)^{1/2} \approx 1 + \frac{1}{2}\epsilon$ to say that the distance is approximately $[1 + \frac{1}{2}h(t)]L$, meaning that an observer with an absolute ruler would say that each mass is oscillating toward or away from the system’s center of mass with an acceleration of $a = \frac{1}{2}\ddot h (\frac{1}{2}L)$ (since each mass is $\frac{1}{2}L$ from the center) = $\frac{1}{4}\ddot h L$.
However we might interpret this in general relativity, from a Newtonian perspective, this acceleration must be caused by a force, meaning that each mass behaves as if it were experiencing a force of magnitude
\be
\label{GW-force}
F=\frac{1}{4}m {\ddot h}(t)L
\ee
toward or away from each the system’s center of mass. If the two masses happened to be connected by a spring, this effective force would measurably compress or expand the spring: indeed, the spring would behave as if it were being compressed by a total force of $F=\frac{1}{2}m {\ddot h}(t)L$.

The above discussion demonstrates that the stretching and compression of space by GWs
do yield observable physical effects.
In the following, we focus on case 2 and provide three extreme examples (destruction of organs, earthquakes, and tides) with which readers unfamiliar with general relativity can 
imagine the physical effects of GWs intuitively in terms of Newtonian mechanics.
As the waveform of a GW, we consider a monochromatic wave given by
\be
h(t)=h_0 \cos (2\pi f_{\rm gw}t).
\ee

\section{Example 1: Organ destruction}
\label{sec:org}

The first example we study is the potential destruction of the human body 
caused by passing GWs.
With the typical realistic strain, the effect of GWs on the human body is totally negligible.
However, as we have discussed in \S\ref{sec:GA}, for a sufficiently large amplitude of GWs, 
in principle, GWs could destroy organs or tissues inside our body.
Let us provide an order-of-magnitude estimation of the stretching of GWs at which 
organs are destroyed 
by determining the force exerted by GWs of various amplitudes and frequencies on the parts of an organ and comparing against the maximum force the organ can tolerate without being destroyed.

In the order-of-magnitude estimation,
we may replace an organ with two massive objects separated by a distance equivalent to 
the size of the organ
and solve Eq.~(\ref{GW-force}) as a force acting on the organ.
Taking $L=10~{\rm cm}$ and $m=0.1~{\rm kg}$ as representative values for an organ (e.g., heart)
and scaling frequency $f_{\rm gw}$ relative to $100~{\rm Hz}$, which is the typical frequency for an event measured by LIGO, Virgo, or KAGRA,
we find that the amplitude of the total compressive force on the organ is approximately given by 
\be
\label{force-organ}
F\approx 10^3~{\rm N} ~{\left( \frac{f_{\rm gw}}{100~{\rm Hz}} \right)}^2~h_0.
\ee
Roughly speaking, an organ will collapse if more than $\sim10^3~{\rm N}$ is exerted on it~\cite{Rosen:2008}.
By adopting this criterion, 
we find that an organ is destroyed if the GW amplitude $h_0$ satisfies
\be
h_0 \gtrsim {\left( \frac{f_{\rm gw}}{100~{\rm Hz}} \right)}^{-2}.
\ee

\begin{figure}[t]
 \begin{center}
   \includegraphics[clip,width=0.99\columnwidth]{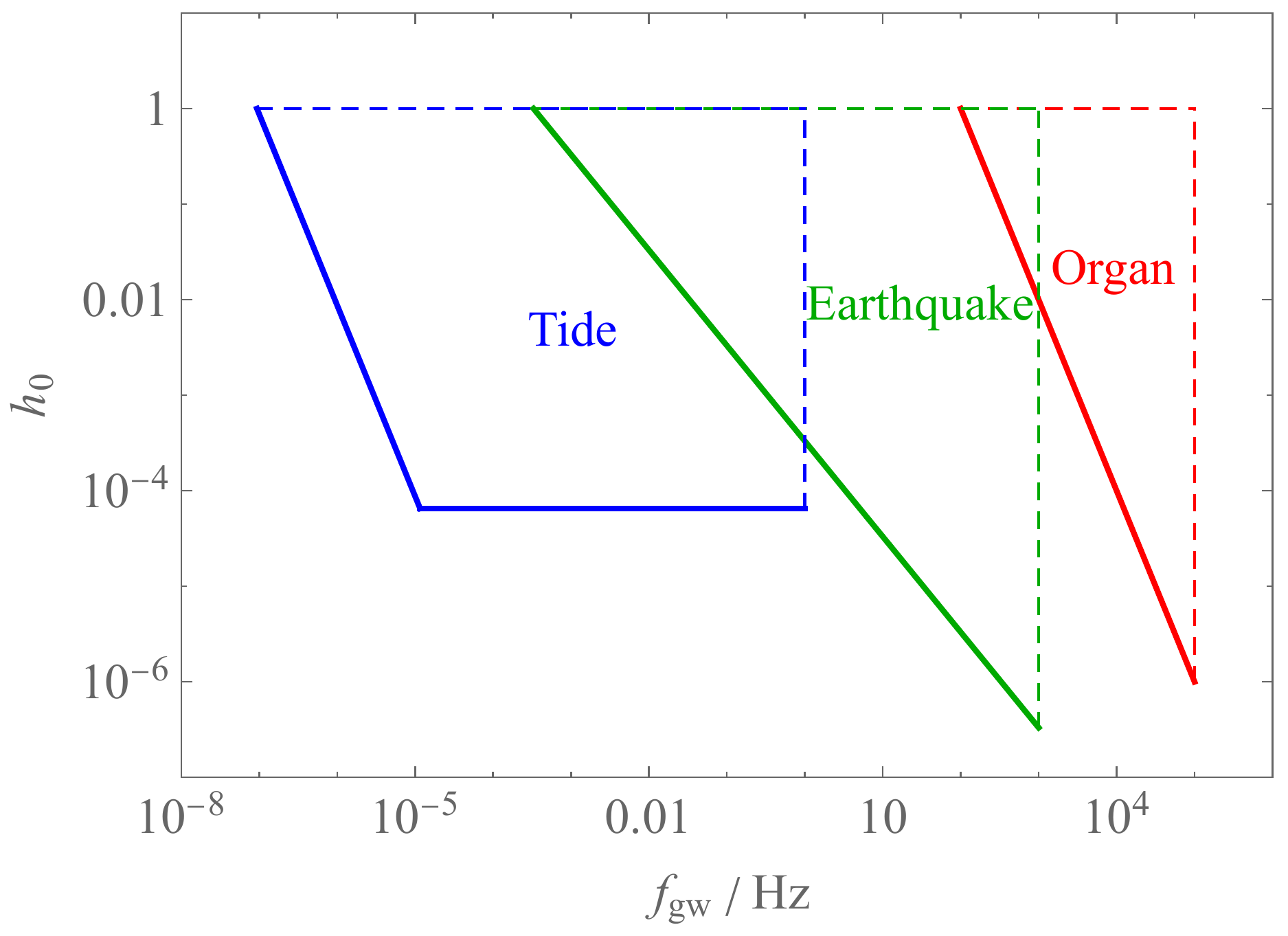}
   \caption{
   GW amplitudes $h_0$ sufficiently strong to cause large effects in terms of frequency $f_{\rm gw}$.
   The solid lines indicate the threshold beyond which human lives would be threatened, 
   whereas the dashed lines indicate the possible boundaries of the validity of the approximations we employed.
   The red, green, and blue lines are, respectively, the thresholds, beyond which the GWs cause organ destruction, major earthquakes, and large tides. 
   }
   \label{constraint-h0}
 \end{center}
\end{figure}

The red solid line in Fig.~\ref{constraint-h0} shows the boundary of this equality 
as a function of $f_{\rm gw}$.
The lower limit of $f_{\rm gw}$ around $10^2~{\rm Hz}$ originates from the criterion $h_0 <1$ for 
which our linear approximation is valid.
Cases where $h_0 >1$ are beyond the scope of this paper.
The upper limit of $f_{\rm gw}$ is the inverse of the time that a sound wave travels 
across an organ:
\be
f_{\rm gw} < \frac{c_{\rm o}}{L}=10^5~{\rm Hz} ~{\left( \frac{c_{\rm o}}{10^3~{\rm m/s}} \right)} 
{\left( \frac{L}{10~{\rm cm}} \right)}^{-1}, \label{organ-adiabatic}
\ee
where $c_{\rm o}$ is the speed of sound passing across the organ.
Above this frequency, the force from GWs changes its direction before it affects the whole organ, 
and it is not clear whether the organ is destroyed even if the force given by Eq.~(\ref{force-organ})
is applied.
Hence, in our analysis, we consider the frequency range that satisfies this condition.
In Fig.~\ref{constraint-h0}, the boundaries $h_0=1$ and $f_{\rm gw}=10^5~{\rm Hz}$ are shown 
as red dashed lines.

\section{Example 2: Earthquakes}
\label{sec:eq}

Next, we consider an earthquake that would be induced by the passage of GWs.
A pioneering work was performed by Dyson, 
who calculated the seismic response of the Earth to the passage of 1~Hz GWs~\cite{Dyson:1997gv}.
Constraints using potential seismic signatures have been extensively studied~\cite{Coughlin:2014sca}.
Dyson's computation heavily relies on tensorial equations
and general relativity, which are beyond the scope of this paper.
Instead, here, we consider a back-of-the-envelope calculation.
Quite remarkably, results of our crude calculation reproduce those of
Dyson's computation within a factor of ${\cal O}(1)$.

An earthquake is simply the vibration of the ground propagating as a sound wave.
When a GW of frequency $f_{\rm gw}$ is passing through the Earth,
a sound wave with the same frequency is induced and propagates.
Let us model 
the motion of a segment of the Earth medium with a harmonic oscillator in which a point mass is attached to a spring of length $L$.
As we have discussed in \S\ref{sec:GA}, GWs induce the acceleration of the harmonic 
oscillator of the order of $\pi^2 f_{\rm gw}^2L h_0$.
From the shape of the plane wave $e^{2\pi i x/\lambda_{\rm e}}$, 
where $\lambda_{\rm e}$ is the wavelength of a transverse seismic wave passing through the Earth,
the phases of the ground motion at different locations are nearly coherent
(i.e., the phase difference is $< 2 \pi$) if the separation is less than $\lambda_{\rm e}$.
In other words, modeling the vibration of the ground as oscillations of the harmonic
oscillator is valid for 
the segment of the Earth medium with a (typical) size $\lesssim \lambda_{\rm e}$.
This means that we should take $L$ to be $\lambda_{\rm e}$.
Then, the acceleration of the ground is estimated as 
\be
\label{earthquake-a}
a =\pi^2 c_{\rm e} f_{\rm gw} h_0,
\ee
where $c_{\rm e} = \lambda_{\rm e} f_{\rm gw}$ is the speed of a transverse seismic wave passing through the Earth, 
which is typically $3\times 10^3$~m/s. 
Note also that since $f_{\rm gw}=c/\lambda_{\rm gw}=c_{\rm e}/\lambda_{\rm e}$ it holds that $\lambda_{\rm e} \ll \lambda_{\rm gw}$.
The acceleration~\eqref{earthquake-a} is the same as that obtained by Dyson up to a factor of ${\cal O}(1)$~\cite{Dyson:1997gv}.

The induced earthquake will be catastrophic if the acceleration is greater than
the gravitational acceleration $g\approx 9.8$~m/s$^2$.
Thus, by equating acceleration (\ref{earthquake-a}) to $g$, we obtain the critical GW
amplitude $h_0$ that yields a serious earthquake:
\be
h_0 =\frac{g}{\pi^2 c_{\rm e} f_{\rm gw}}.
\ee

The green solid line in Fig.~\ref{constraint-h0} shows $h_0$ 
that causes an earthquake with ground acceleration 
comparable to the gravitational acceleration.
As in \cite{Dyson:1997gv}, 
we focus on the GW frequency range $1~{\rm mHz}< f_{\rm gw} < 1~{\rm kHz}$. 
In this range,
it is a good approximation to treat the surface of the Earth as a flat plane since
the wavelength of the induced seismic waves is much smaller than the size of the Earth
and larger than the surface irregularities.

\section{Example 3: Tides}
\label{sec:tid}

The third effect we consider is tides, i.e., the rise and fall of sea levels, which
are usually generated by the gravitational forces from the Moon and Sun.
Similarly, when GWs pass through the Earth, they can generate tides.
Tides involve very complicated processes and it is difficult to analyze
those generated by the passage of GWs.
As in the previous two examples, 
we will evaluate tides by a crude approximation.
To this end, we treat the Earth as a rock sphere completely encompassed 
by a water shell of uniform depth $H$. 
In this model, we assume that the rock sphere is so rigid that it is not deformed 
by GWs and only the water shell is freely deformed (since it is liquid) 
by the force given by Eq.~(\ref{GW-force}).

For this model, the first task is to determine the length scale $L$.
We make a rough estimate;
if the period of GWs is longer than one day,
the ocean on the global scale will coherently respond to GWs similarly to
the case in response to the tidal force generated by the Moon's gravitational force.
In this case, the natural scale of $L$ will be the Earth radius $R_\oplus$.
Assuming $L=R_\oplus$, 
the force from the GWs exerted on the volume element $\Delta V$ is given by
\be
\label{gw-tide}
F\simeq 
\pi^2 f_{\rm gw}^2 \rho_{\rm w}R_\oplus h_0 \Delta V,
\ee
where $\rho_{\rm w}$ is the density of water.
Conversely, if the period of GWs is much shorter than one day
but longer than the time scale of the sound wave traveling across the depth $H$,
the response of the sea level will not be the same as that in the low-frequency case.

To understand this situation, let us consider the motion of a harmonic
oscillator with a periodic external force as a simple model:
$m{\ddot x}=-kx +m \omega^2 h_0 L \sin (\omega t)$.
Here, $x$ and the external force are assumed to represent 
the change in sea level and the effect by GWs, respectively.
Note that we have taken the coefficient of the external force to be $m \omega^2 h_0 L$;
with this expression, $h_0 L$ does not depend on $\omega$ and
is simply proportional to $h_0$ [see Eq.~(\ref{GW-force})].
Assuming $x=x_0 \sin (\omega t)$, $x_0$ is given by
\be
x_0=\frac{h_0 L \omega^2}{\omega_0^2-\omega^2},
\ee
where $\omega_0=\sqrt{k/m}$.
In the low-frequency limit $\omega \ll \omega_0$, we have $x_0 = h_0 L \omega^2/\omega_0^2$, namely,
$x_0$ is determined by the balance between the term $-kx$ 
and the external force.
On the other hand, in the high-frequency limit $\omega \gg \omega_0$, we have $x_0 =-h_0 L$.
Apparently, this might look inconsistent with the standard result of the forced harmonic oscillator, for which the amplitude goes to zero in the high-frequency limit.
The point is that, as stressed above, the GW driving force actually increases with frequency for a given GW amplitude $h_0$, and this leads to the conclusion that in the high-frequency limit $x_0$ becomes a constant value $-h_0 L$ independent of the frequency of the external force $\omega$.
Given these observations,
it would be reasonable to consider that the tide induced by GWs becomes
independent of $f_{\rm gw}$ for $f_{\rm gw} \gg f_{\rm day}=1/{\rm day}$.

The next nontrivial task is to estimate the magnitude of the tide
induced by the force given in Eq.~\eqref{gw-tide}, which, in principle, can be done by solving the shallow-water equation.
However, even at the level of the order-of-magnitude estimate, 
how this can be achieved is nontrivial.
To overcome this issue, we simply compare force (\ref{gw-tide}) with the
tidal force induced by the Moon.
For $f_{\rm gw} < f_{\rm day}$,
we assume that the level of the tide induced by GWs is the same as 
that of an ordinary tide
if the two forces have the same magnitude.
The tidal force from the Moon acting on the volume element $\Delta V$
is estimated as
\be
\label{tide-moon}
F_{\rm L} \simeq \frac{GM_{\rm L} \rho_{\rm w}\Delta V}{D^3}R_\oplus.
\ee
Here, $M_{\rm L}$ is the lunar mass, and $D$ is the lunar distance.
By equating Eq.~(\ref{gw-tide}) to Eq.~(\ref{tide-moon}), we obtain
the minimum $h_0$ that yields a tide comparable to the ordinary one as
\be \label{h0-tide}
h_0 =\frac{GM_{\rm L}}{\pi^2 D^3 f_{\rm gw}^2}.
\ee
On the other hand, for $f_{\rm gw} \gg f_{\rm day}$, 
on the basis of the above argument on the forced oscillator,
we assume that the critical $h_0$ is given
by the right-hand side of Eq.~\eqref{h0-tide} with $f_{\rm gw}$ being 
replaced by $f_{\rm day}$.

The blue solid line in Fig.~\ref{constraint-h0} shows $h_0$ that causes a tide comparable
to that caused by the Moon.
The increase in sea level to this amount will threaten 
the lives of many people residing in coastal regions.
The upper limit for $f_{\rm gw} \lesssim 1~{\rm Hz}$ is imposed by the condition for the validity of the shallow-water approximation
that the time for sound to pass across the depth of the ocean is shorter than the GW frequency.

\section{GWs from BH binaries}
\label{sec:bh}

In the previous sections, we considered extreme situations where the GWs cause visible effects
on humans by causing stretching and shrinking of space, 
and we obtained the parameter regions in the amplitude--frequency space, 
as shown in Fig.~\ref{constraint-h0}.
Now it is interesting to consider converting the parameter regions for the GWs to those 
for BH binaries.
The strongest GWs are emitted during the last moment of the merging of BHs
in a binary system.
Therefore, it is natural to try to relate GWs to BH binary systems in such a way.

We assume that a BH binary system consists of two equal-mass BHs each of which has mass $M$
and is located at the distance $r$ from the Earth.
Precise modeling of the GW waveform from such a phase requires numerous computations
based on general relativity.
Since we are working at the level of the order-of-magnitude estimate,
we do not use the results obtained from full computations but
make a rough estimate of the GW amplitude and frequency as follows.

At the last inspiral phase, the distance between BHs is
comparable to the size of the BHs, i.e., the Schwarzschild radius $r_{\rm s}=2GM/c^2$,
and the BHs are orbiting at a relativistic speed. 
In this extreme situation, the disturbance of the gravitational field caused by the motion of the BHs, which propagates outward as GWs, would be comparable to the gravitational field generated by the BHs when they are not moving. 
Thus, it is natural to think that the GW amplitude
around the BHs is also ${\cal O}(1)$. 
The GW amplitude attenuates inversely proportionally to the propagation distance.
Thus, by denoting $r$ as the distance to the GW source,
we find that the GW amplitude $h_0$ on the Earth would be on the order of
\be
h_0 \simeq \frac{2GM}{c^2r}.
\ee
The period of GWs is given by the time scale of the change of the 
GW source, which is $r_{\rm s}/c$.
Therefore, the frequency of the GWs would be on the order of
\be
\label{fgw-M}
f_{\rm gw} \simeq \frac{c}{r_{\rm s}}=\frac{c^3}{2GM}.
\ee
This clearly shows that there is a one-to-one correspondence between
$f_{\rm gw}$ and $M$.

\begin{figure}[t]
 \begin{center}
   \includegraphics[clip,width=0.99\columnwidth]{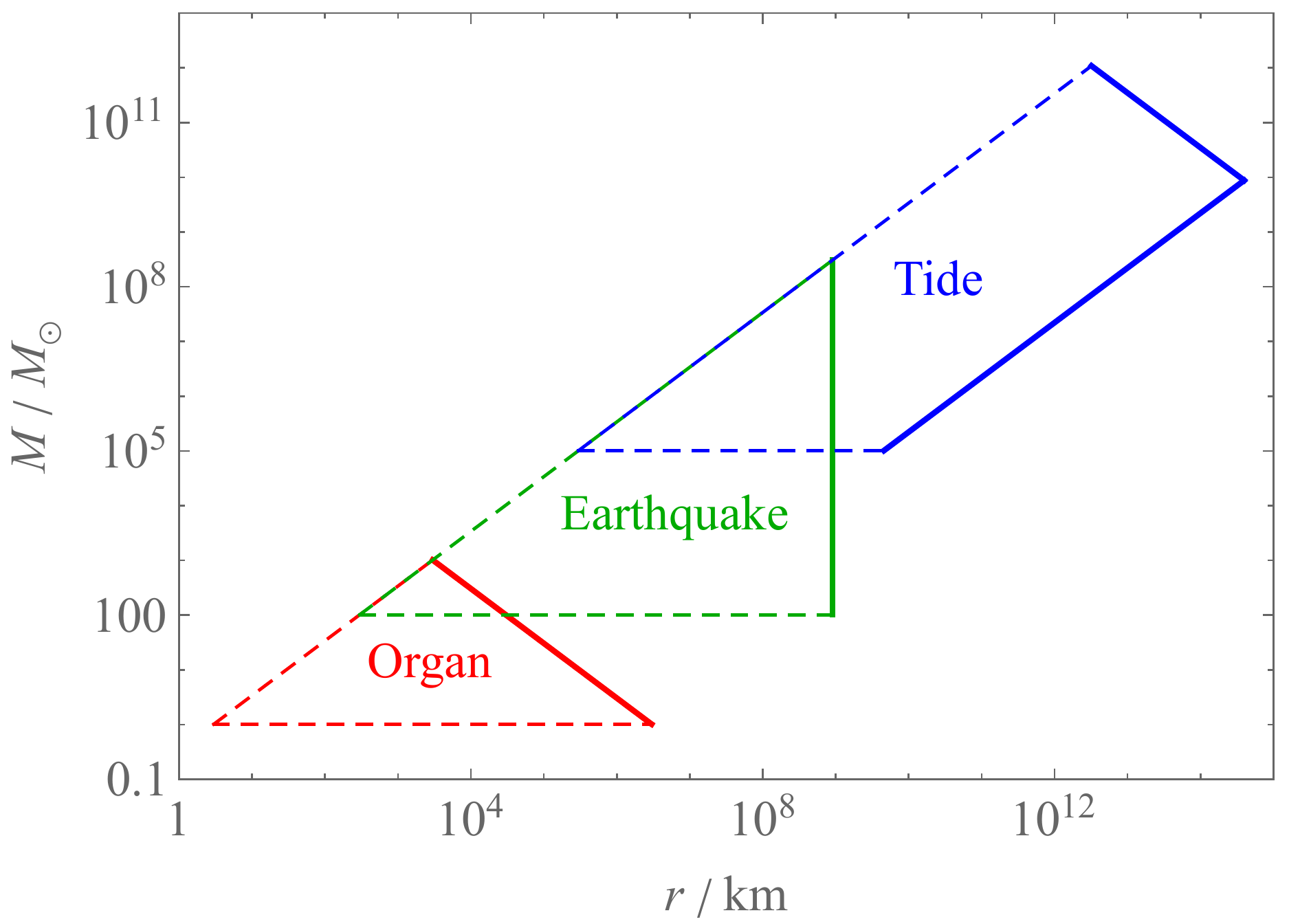}
   \caption{
   Parameter region corresponding to Fig.~\ref{constraint-h0} in terms of the BH mass $M$ and the distance $r$ from the Earth assuming that GWs are emitted at the last inspiral stage of two BHs of equal mass. The BH mass $M$ is normalized by the solar mass $M_\odot \approx 1.99\times 10^{30}$kg. 
   }
   \label{constraint-h0-BH}
 \end{center}
\end{figure}

In the previous sections, we estimated the critical values of $h_0$
above which phenomena catastrophic to humans occur at various values 
of $f_{\rm gw}$. 
By using Eq.~(\ref{fgw-M}) to replace $f_{\rm gw}$ with $M$, 
we can draw curves corresponding to the 
critical values of $h_0$ in the $M$--$r$ plane, as shown in Fig.~\ref{constraint-h0-BH}.
As expected, the results in Fig.~\ref{constraint-h0-BH} suggest 
that the GWs would cause serious damage 
only if too-heavy BHs merged at too-close distances from the Earth.
It would be a good exercise to think about the farthest and/or lightest parameter value for each case.
For instance, let us consider the farthest parameter value 
$(r,M)\sim (10^{15}{\rm km},10^{10}M_\odot)$ for the tide.
It means that a large tide would be induced by GWs if two supermassive BHs each with a 
mass of $\sim 10^{10}M_\odot$ merged at about 100 light years from the Earth.
To get a sense of this situation, let us compare the distance and the mass $(r,M)\sim (10^{15}{\rm km},10^{10}M_\odot)$ of a BH
with those of Sgr~A$^*$, which is the supermassive BH at the center of the Milky Way galaxy.
The mass of Sgr~A$^*$ is $4.2\times 10^6M_\odot$ and the distance is $2.7\times 10^4$ 
light years from the Earth~\cite{Abuter2019}.
Therefore, even the farthest parameter value means that too-heavy BHs merged at 
too-close distances from the Earth.
If such supermassive BHs existed, the Solar System would be destroyed by their gravitational force far before they merge.
Therefore, the earthquakes and tides generated by GWs are totally irrelevant to reality. 
However, when space travel to a BH binary systems becomes popular in the future, 
the effect of GWs on organs may be relevant during such travel.
In that case, the red boundary in Fig.~\ref{constraint-h0-BH} must be included 
as a caution in the flight plan for astronauts.

\section{Conclusion}
\label{sec:con}

GWs are ripples of spacetime traveling across the universe.
Not only spacetime but also everything in nature are stretched and compressed by GWs.
Therefore, in principle, they could cause some large effects.
After clarifying the conceptual issue about the detectability of GWs and
translating the effects of GWs into the language of Newtonian mechanics,
we performed the order-of-magnitude estimation for visible phenomena, such as organ destruction, earthquakes, and tides, that would be caused by large-amplitude GWs.
We obtained the amplitude--frequency space in which GWs would cause some large effects.
As a reference, we converted the parameter regions to those in BH mass--distance space.
These are interesting examples highlighting the nature of GWs
and are useful for gaining some understanding of the theory of relativity.

\acknowledgments

H.M.\ was supported by Japan Society for the Promotion of Science (JSPS) Grants-in-Aid for Scientific Research (KAKENHI) Grants No.\ JP18K13565 and No.\ JP22K03639.
T.S.\ was supported by the MEXT Grant-in-Aid for Scientific Research on Innovative Areas No.\ 17H06359, No.\ 19K03864, and No.\ 21H05453.

\bibliographystyle{JHEPmod}
\bibliography{refs}

\providecommand{\href}[2]{#2}\begingroup\raggedright\begin{thebibliography}{10}

\bibitem{Abbott:2016blz}
{\scshape LIGO Scientific, Virgo} collaboration, B.~Abbott et~al., {\it
  {Observation of Gravitational Waves from a Binary Black Hole Merger}},
  \href{https://doi.org/10.1103/PhysRevLett.116.061102}{Phys. Rev. Lett.
  {\bfseries 116} (2016) 061102}
  [\href{http://arxiv.org/abs/1602.03837}{{\ttfamily arXiv:1602.03837}}].

\bibitem{LIGOpress}
\url{http://mediaassets.caltech.edu/gwave}.

\bibitem{Schutz:2021xns}
B.~F. Schutz, {\it {Intuition in Einsteinian Physics}},
  \href{http://arxiv.org/abs/2106.01820}{{\ttfamily arXiv:2106.01820}}.

\bibitem{Cervantes-Cota:2016zjc}
J.~L. Cervantes-Cota, S.~Galindo-Uribarri and G.-F. Smoot, {\it {A Brief
  History of Gravitational Waves}},
  \href{https://doi.org/10.3390/universe2030022}{Universe {\bfseries 2} (2016)
  22} [\href{http://arxiv.org/abs/1609.09400}{{\ttfamily arXiv:1609.09400}}].

\bibitem{Misner:1973prb}
C.~W. Misner, K.~S. Thorne and J.~A. Wheeler, {Gravitation}. W. H. Freeman, San
  Francisco, 1973.

\bibitem{Creighton}
J.~D.~E. Creighton and W.~G. Anderson, Gravitational-wave physics and
  astronomy: An introduction to theory, experiment and data analysis, Wiley
  Series in Cosmology. Wiley-VCH, 2011.

\bibitem{Maggiore:2007ulw}
M.~Maggiore, {Gravitational Waves. Vol. 1: Theory and Experiments}, Oxford
  Master Series in Physics. Oxford University Press, 2007.

\bibitem{Saulson:1997ck}
P.~R. Saulson, {\it {If light waves are stretched by gravitational waves, how
  can we use light as a ruler to detect gravitational waves?}},
  \href{https://doi.org/10.1119/1.18578}{Am. J. Phys. {\bfseries 65} (1997)
  501}.

\bibitem{Schutz:1984nf}
B.~F. Schutz, {\it {Gravitational waves on the back of an envelope}},
  \href{https://doi.org/10.1119/1.13627}{Am. J. Phys. {\bfseries 52} (1984)
  412}.

\bibitem{Farr:2011cw}
B.~Farr, G.~Schelbert and L.~Trouille, {\it {Gravitational Wave Science in the
  High School Classroom}}, \href{https://doi.org/10.1119/1.4738365}{Am. J.
  Phys. {\bfseries 80} (2012) 898}
  [\href{http://arxiv.org/abs/1109.3720}{{\ttfamily arXiv:1109.3720}}].

\bibitem{Burko:2016vnu}
L.~M. Burko, {\it {Gravitational Wave Detection in the Introductory Lab}},
  \href{https://doi.org/10.1119/1.4981036}{Phys. Teacher {\bfseries 55} (2017)
  288} [\href{http://arxiv.org/abs/1602.04666}{{\ttfamily arXiv:1602.04666}}].

\bibitem{Mathur:2016cox}
H.~Mathur, K.~Brown and A.~Lowenstein, {\it {An analysis of the LIGO discovery
  based on Introductory Physics}}, \href{https://doi.org/10.1119/1.4985727}{Am.
  J. Phys. {\bfseries 85} (2017) 676}
  [\href{http://arxiv.org/abs/1609.09349}{{\ttfamily arXiv:1609.09349}}].

\bibitem{Hilborn:2017liy}
R.~C. Hilborn, {\it {Gravitational waves from orbiting binaries without general
  relativity}}, \href{https://doi.org/10.1119/1.5020984}{Am. J. Phys.
  {\bfseries 86} (2018) 186} [\href{http://arxiv.org/abs/1710.04635}{{\ttfamily
  arXiv:1710.04635}}].

\bibitem{Rosen:2008}
J.~Rosen, J.~Brown, S.~De, M.~Sinanan and B.~Hannaford, {\it {Biomechanical
  properties of abdominal organs in vivo and postmortem under compression
  loads}}, \href{https://doi.org/10.1115/1.2898712}{Journal of Biomechanical
  Engineering {\bfseries 130} (2008) 021020}.

\bibitem{Dyson:1997gv}
F.~Dyson, {\it {Seismic response of the earth to a gravitational wave in the
  1-Hz band}}, \href{https://doi.org/10.1086/149986}{Astrophys. J. {\bfseries
  156} (1969) 529}.

\bibitem{Coughlin:2014sca}
M.~Coughlin and J.~Harms, {\it {Upper Limit on a Stochastic Background of
  Gravitational Waves from Seismic Measurements in the Range 0.05\textendash{}1
  Hz}}, \href{https://doi.org/10.1103/PhysRevLett.112.101102}{Phys. Rev. Lett.
  {\bfseries 112} (2014) 101102}
  [\href{http://arxiv.org/abs/1401.3028}{{\ttfamily arXiv:1401.3028}}].

\bibitem{Abuter2019}
{\scshape GRAVITY} collaboration, R.~Abuter et~al., {\it {A geometric distance
  measurement to the Galactic Center black hole with 0.3\% uncertainty}},
  \href{https://doi.org/10.1051/0004-6361/201935656}{Astron. Astrophys.
  {\bfseries 625} (2019) L10}
  [\href{http://arxiv.org/abs/1904.05721}{{\ttfamily arXiv:1904.05721}}].

\end{thebibliography}\endgroup

\end{document}